\documentclass[a4paper]{article}

\usepackage{amsmath}
\usepackage{graphicx}

\begin{document}


\title{Noise level forecasts at 8-20MHz and their use for morphological
 studies of ionospheric absorption variations at EKB ISTP SB RAS radar}

\author{Oleg I.Berngardt}

\maketitle

\begin{abstract}
In this paper, a method is described for using 8-20MHz noise 
absorption effect for realtime detecting radiowave absorption periods.
The method is based on two empirical autoregression models of noise dynamics.
The first (rough) prediction model is based on radar measurements
of daily minimal noise dynamics averaged over 28-days with specially
calculated weight coefficients. The second (fine) prediction model uses real-time
scaling of rough model. The scaling is based on the comparison of this model with the experimental noise
observations during previous 5 days. The models
are based on the whole EKB ISTP SB RAS radar dataset (2013-2018). 
The rough model allows one to estimate the boundary beyond which the noise variations can 
be associated with absorption periods with a high degree of certainty.
A joint analysis of simultaneous data on neighboring radar beams and
at several frequencies reduces the detection
errors, and allows to identify absorption events with a higher degree of confidence.
Use of fine model allows to estimate absorption.
 The technique is validated by frequency dependence
of absorption during two-frequency measurements. The
found frequency dependence has an average exponent of the order of
-1.5, which is in good agreement with the literature data and the data
obtained earlier in analysis of solar
X-ray flares. 
The use of the detection technique at EKB radar shows that most probable absorption 
over absorption events is about -0.65dB. Analysis of absorption of different amplitudes 
shows that low-intensity absorption events (0..-1.3dB) have slight local time dependence 
and mostly observed at north directions. For the storng absorption events (stronger than -1.3dB)
the local time dynamics correlates well with noise level dynamics, and usually fills the whole radar 
field-of-view.
\end{abstract}




\section{Introduction}

The use of 8-20MHz radio noise, measured by short-wave radars for
the diagnosis of the lower part of the ionosphere, is an intensively
developed method \cite{BERNGARDT_2018,Bland_2018,Bland_2019,Berngadt_2019,Chakraborty_2019}.
The physical processes that cause this effect \cite{Berngadt_2019}
are close to the physical processes that cause the absorption of the
signals scattered from the earth\textquoteright s surface (ground
scatter, GS) \cite{Watanabe_2014,Fiori_2018}, so GS signals and noise
level both can be used to study the absorption of radiwaves \cite{Chakraborty_2019}.
The first observations of the absorption of 8-20~MHz radio noise 
during X-ray solar flares at mid-latitude radars \cite{BERNGARDT_2018} showed
the efficiency of this method. Subsequent statistical analysis
according to the data of world-wide radar network showed usefulness
of this method in the lit time: it allows approve the main influence
of the ionosphere near the radar to the absorption; it allows demonstrate the
contributions of D- and E-layers to the absorption; it allows estimate the
frequency dependence of absorption; it allows localize the region
in which the absorption measured \cite{Berngadt_2019}. Studies of absorption
associated with the corpuscular solar radiation allow demonstrate
a good agreement between radio noise 8-20MHz absorption and standard
observations by riometers \cite{Bland_2018,Bland_2019}, allowing
to prove the correctness of the technique in these cases too.

Therefore, an important practical issue is creation an automatic detector system 
for detecting the periods of radio signal
absorption from the 8-20MHz noise data. Similar systems are usually based
on riometric noise prediction model, allowing to make an accurate forecast of the noise
level for a significant period of time ahead of the observations,
exceeding the duration of the expected effects. The prediction
time and accuracy of the forecast determines the accuracy of absorption detection technique. It should be noted that
the periods of radio noise absorption can be short-term (up
to several hours)\cite{BERNGARDT_2018,Berngadt_2019} or long-term
\cite{Bland_2018,Bland_2019}. Most of the recent papers is currently devoted
to case studying during selected events  \cite{BERNGARDT_2018,Bland_2018,Bland_2019,Berngadt_2019},
and a complete analysis of all cases of absorption using coherent radars has not yet been
carried out.

\begin{figure}
\includegraphics[scale=0.4]{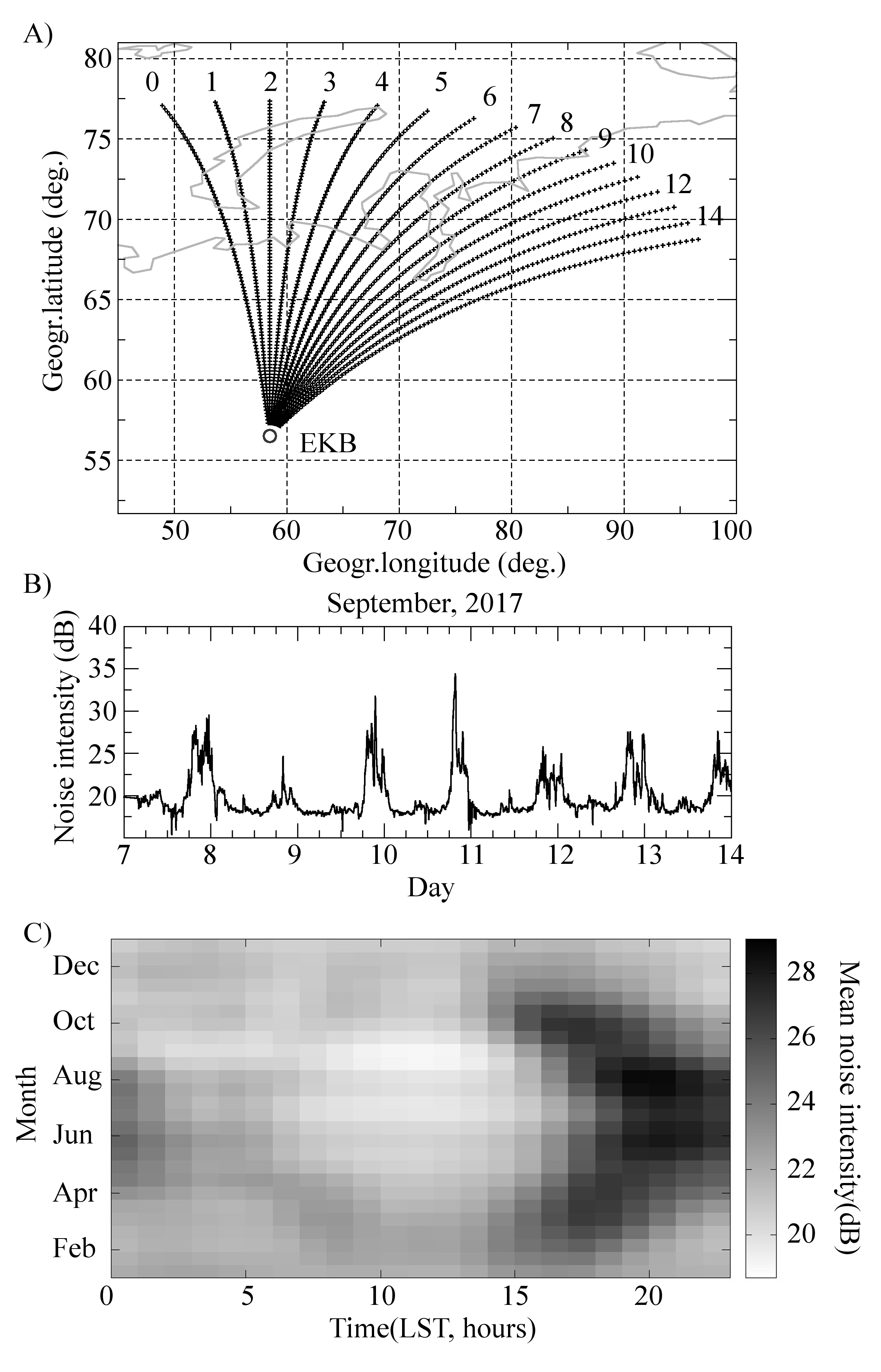}
\caption{A) Geographical location of the EKB ISTP SB RAS radar and its field
of view. The lines correspond to the trajectories (beams) of the radiosignal
propagation, the numbers mark the beam numbers. B) An example
of noise level variations during September 7-13, 2017; C) The average
noise level on beam 2 as a function of the month number and local
solar time(LST). }
\label{fig:Fig1}
\end{figure}

The EKB ISTP SB RAS radar is a mid-latitude decameter radar of CUTLASS
kind \cite{Lester_2004} developed at the University of Leicester,
similar in design and software to SuperDARN (Super Dual Auroral Radar
Network) \cite{Greenwald_1995,Chisham_2007,Nishitani_2019}
radars. The radar was bought under the financial support of SB RAS,
and was installed together with the IGF UrB RAS near Ekaterinburg,
in Arti observatory, IGF UrB RAS (56.5N,
58.5E). The radar field of view is about 1 million square kilometers
and is shown in Fig.\ref{fig:Fig1}A. In regular mode, the
radar operates in 16 discrete directions (beams) with  1-2 minute temporal
resolution, with 45 km spatial resolution, and
at the operating frequencies 8-20 MHz. The radar transmits sequences
of short sounding pulses and processes the received signals, scattered
from surface and ionospheric irregularities. This information is used
to investigate an ionospheric characteristics within the radar field-of-view.
Between the soundings, the radar measures the noise level near the sounding
frequency to select the frequency at which
the background noise level is minimal. An example of the noise dynamics,
measured at a single beam and a single frequency is shown in Fig.\ref{fig:Fig1}B.
The radar operates simultaneously in two frequency channels,
which makes it possible to measure the properties of the scattered
signal and the noise level in two frequencies simultaneously
and independently.

It was shown that the analysis of the minimal noise level allows using
the radar as a mid-latitude riometer\cite{BERNGARDT_2018,Bland_2018,Bland_2019,Berngadt_2019}.
An essential issue associated with this use is the reliable prediction
of noise dynamics to calculate the amplitude of noise attenuations 
considered as absorption.

The short-wave radio noise is produced by a superposition
of signals from various sources (anthropogenic, space and atmospheric), and each of
them has its own seasonal, daily and geographical dependency
\cite{ITU_R}. The mechanism of long-range multihop radiwave
propagation at 8-20 MHz makes the prediction of the noise
level a difficult task. The radio noise observed at SuperDARN and
similar radars has significant daily and seasonal dependencies \cite[fig.6]{Ponomarenko_2016}
and currently its reference models suitable for reliable forecast
are apparently absent. The average daily-seasonal noise level dynamics 
at EKB ISTP SB RAS radar is shown in Fig.\ref{fig:Fig1}C for the
beam 2, directed to the North. It can be seen from the figure that
the mean daily noise dynamics significantly depends on the season
and the noise level has a pronounced maximum during sunset solar terminator
and a minimum in the local noon. This effect is similar
to the effect observed earlier by the Canadian SAS SuperDARN radar \cite[fig.6]{Ponomarenko_2016}.
Therefore, the use of simple stationary models for noise prediction
is problematic, and the predictive model should either include the
physical mechanisms of the noise formation or use the noise
historical observations.

\section{Noise prediction model}

The traditional approach used in riometric observations is to use the
simplest mean daily noise model, averaged over previous
30 days \cite{Heisler_1967}, or use database of quiet days to select 
the quiet day curve \cite{Canopus}. We will use the first one as the simplest.
Its use can be justified in cases
where the effects of the propagation of radio waves can be neglected
(which is acceptable for the frequencies 30-40 MHz used by riometers).
In spite the possibility and effectiveness of using such models for
noise prediction in some cases at SuperDARN radar \cite{Bland_2018}, 
their validity and
accuracy for all cases is not obvious. Due to the fact that the noise
level on radars is affected not only by the absorption itself, but
also by the ionospheric F-layer dynamics affecting
the propagation trajectory of the noise \cite{Berngadt_2019}, the use of
the mean daily noise model is not entirely justified.

In this paper, the problem of predicting 8-20 MHz radio noise level
from experimental data is considered as a combination of two problems
- making a rough prediction model of the minimal noise level (i.e.
the long-term prediction of dynamics of minimal noise  associated with
regular processes both in the ionosphere and in the noise sources)
and making a technique for forecasting mean noise level in real-time 
by correcting this model. The accuracy of the 
forecast will depend on a combination
of the accuracy of these two techniques. In the both cases, we used an 
autoregression analysis to build the optimal prediction algorithm.

To build the first (rough) prediction model a regression coefficients
were found (based on a large observational base of EKB radar measurements during
2013-2016), that optimally predict the minimal noise level based on the measurements of the minimal noise in the same local time over previous
days. The prediction model is built as an autoregression with the calculated
coefficients. The minimal noise is used because the radar often detects
interference from a nearby ionosonde and
other anthropogenic sources, and the use of minimal noise should compensate
this interference.

To build the second (fine) prediction model, we calculated coefficients
allowed us to scale (calibrate) the rough model of minimal noise to
optimally fit the experimental data 12 hours ahead. The coefficients
were obtained using the entire history over the several previous days
and using the rough model
of minimal noise for all these moments. The analysis of the whole EKB 
dataset  make it possible
to determine the optimal duration of the calibration period, and the
values of the optimal weighting coefficients for the calibration.

Let us consider these two models in details.

\paragraph{Rough prediction model of the minimal noise level}

To build the model of the diurnal dynamics of the minimal noise, the
approximate 24-hour periodicity of the noise level is taken into account 
(shown in Fig.\ref{fig:Fig1}B). 
In spite of a strong variations of noise level  in the experimental data, the amplitude
and shape of the mean diurnal variation of noise level looks changing slowly.
Therefore, as a model of the diurnal noise level dynamics, we use
an autoregression dependence over measurements in the previous days,
at the same time. As mentioned above, a close approach is used for
analysis of a riometric data by using 30-day history to predict mean
daily noise dynamics \cite{Heisler_1967}.

To exclude the interference associated with the operation of the
nearby ionosonde, as well as with iregular nearby anthropogenic
sources, we use the minimal noise levels determined
over the 5-minute period (it corresponds to approximately three scans
over the entire radar field-of-view) as inputs. The 
minimal noise level $f_{m,min}(t)$ for 24 hours ahead is predicted
as the autoregression:

\begin{equation}
\begin{array}{l}
f_{m,min}(t;f_{sound},B_{m})=\sum_{n=1}^{\infty}R_{n}f_{e,min}(t-n\triangle T;f_{sound},B_{m})\\
f_{e,min}(\tau;f_{sound},B_{m})=\left.Min\left(f_{e}(t';f_{sound},B_{m})\right)\right|{}_{t'=\tau-dT/2}^{\tau+dT/2}
\end{array}
\label{eq:model_daily}
\end{equation}
here $dT$ is the time interval of noise minimization (5 minutes),$\triangle T$
is the main period of the diurnal variations (24 hours), $f_{e}(t')$
is the experimentally measured noise level (dB) at the moment
$t'$; $f_{sound}$ is approximate value of sounding frequency (with accuracy
$\pm500kHz$), $B_{m}$ is the radar beam number, $f_{e,min}(\tau;f_{0},B_{m})$
is minimal noise level over  period $[\tau-dT/2,\tau+dT/2]$.

To determine and verify the regression coefficients $R_{n}$,
the whole available EKB radar dataset is divided into two periods - the training
dataset 2013-2016 and the test dataset 2017-2018. The optimal regression
coefficients $R_{n}$, which provide the most accurate prediction
of the minimal noise level, are calculated over the training dataset
2013-2016 as a result of the following minimization:

\begin{equation}
\varOmega_{1}(f_{0},B_{m})=\int\left(f_{m,min}(t;f_{0},B_{m})-f_{e,min}(t;f_{0},B_{m})\right)^{2}dt=min\label{eq:Regr_criteria}
\end{equation}

\begin{figure}
\includegraphics[scale=0.8]{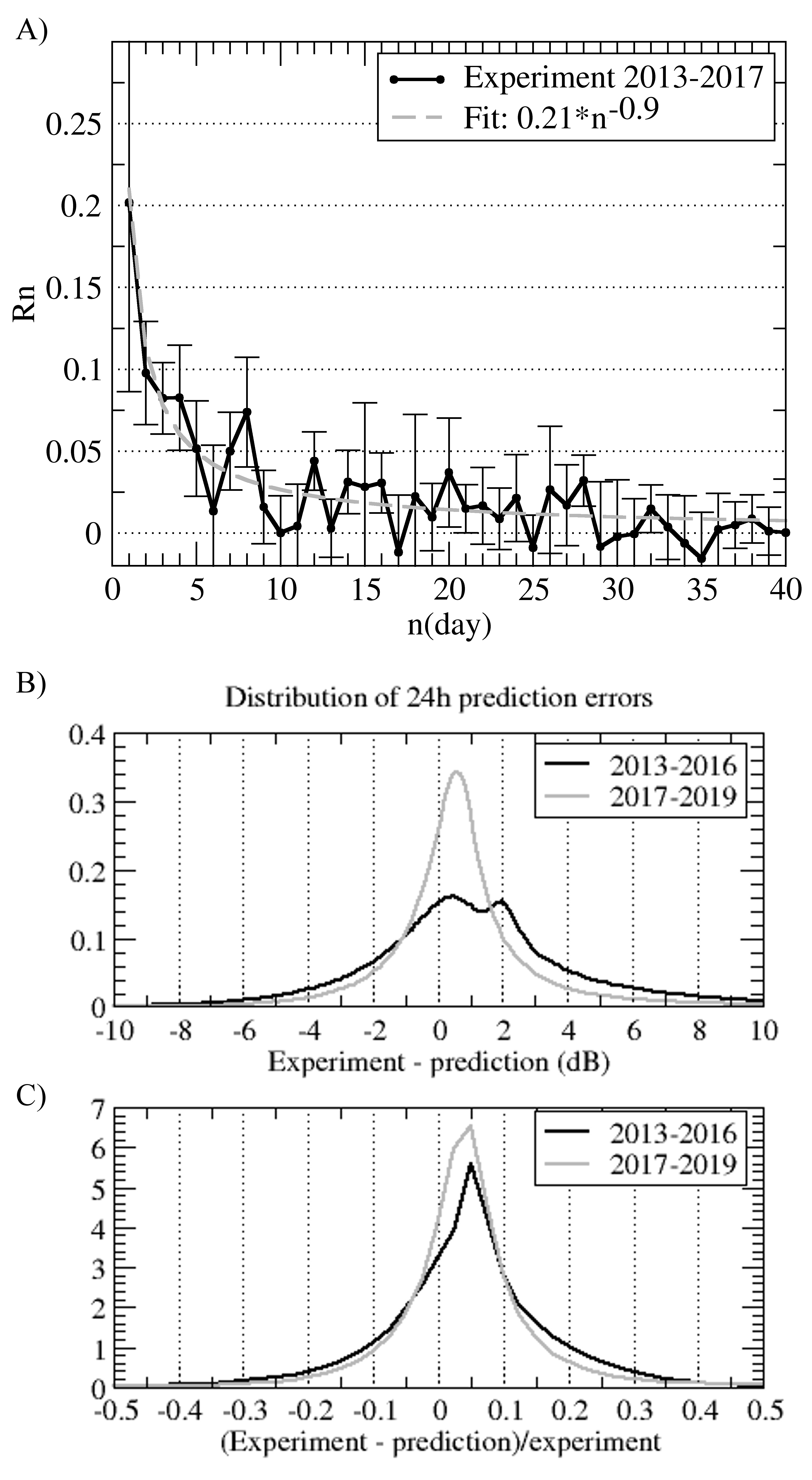}
\caption{A) Weighting coefficients $R_{n}$ for the rough prediction
model (Eq.(\ref{eq:model_daily},\ref{eq:RnModel})) calculated over the training dataset 2013-2016. The
thick solid line corresponds to the averaged experimental $R_{n}$,
thin line corresponds to RMS error of $R_{n}$. The smooth approximation
of $R_{n}$ is shown by the green dashed line. B-C) Model (Eq.(\ref{eq:model_daily},\ref{eq:RnModel})) prediction
accuracy : B) - the distribution
of absolute prediction errors; C) - distribution of relative prediction
errors. The black line in B-C is the distribution of errors over the
training dataset (2013-2016), the gray line is the distribution of
errors over the test dataset (2017-2018).}
\label{fig:Fig2}
\end{figure}

To speed up the calculations, the regression coefficients $R_{n}$ are estimated
over the sequences with 15 minute temporal resolution, and only over
the sequences lasting at least 40 days (sufficient for determining 40 daily regression
coefficients $R_{n}$). The regression is carried out separately
and independently for each beam and in each sounding frequency range.
After this the obtained regression coefficients for all beams and all frequencies
are averaged to obtain the average regression coefficients $R_{n}$
and to estimate the variance of each coefficient. To improve the accuracy
for calculation of Eq.(\ref{eq:Regr_criteria}) only the data from the beams
and the frequencies were used with number of available measurements exceeded
1000. The obtained average regression coefficients $R_{n}$ and their
root mean square (RMS) error are shown in Fig.\ref{fig:Fig2}A by thick
and thin black lines correspondingly.

~

From Fig.\ref{fig:Fig2}A we conclude that the effective
averaging period for the rough predictive model of the noise 
is about 28 days, and the contribution of recent days is
greater than the contribution of older ones. Fig.\ref{fig:Fig2}A shows that the contribution of days older than 28-th is highly variable in sign, so we can neglect them
and limit the calculations by 28-day history. The 28 days averaging
period is in good agreement with the 30-day averaging period used
for prediction of the riometric absorption at higher frequencies 
\cite{Heisler_1967}.
Stronger dynamics of the noise level at frequencies of 8\textendash 20
MHz, associated with background ionospheric dynamics, looks effectively
reducing the averaging range and increases the contribution of recent 
days compared to an older ones into predicted value (Fig.\ref{fig:Fig2}A).

To simplify the calculation of the coefficients, we approximated the
dependence of the regression coefficients $R_{n}$ on the day number
$n$ by the analytical function:

\begin{equation}
R_{n} \approx \left\{ \begin{array}{l}
0.21\cdot n^{-0.9},\,n\in[1..28]\\
0,\,n>28
\end{array}\right.\label{eq:RnModel}
\end{equation}

The approximation is shown in Fig.\ref{fig:Fig2}A by dashed
line. Using the constructed smoothed regression coefficients $R_{n}$
(Eq.(\ref{eq:RnModel})),
the minimal noise level forecast is constructed for a day ahead,
separately for the training dataset 2013-2016, and for the test dataset
2017-2018, and they are compared with experimental observations.
The Fig.\ref{fig:Fig2}B-C show the distribution of prediction errors
for the day ahead according to the model Eq.(\ref{eq:model_daily},\ref{eq:RnModel})
- for absolute errors (Fig.\ref{fig:Fig2}B) and for relative errors (Fig.\ref{fig:Fig2}C).

In Fig.\ref{fig:Fig2}B-C shown that the average absolute
and relative errors are biased for both the training and test datasets.
Narrower error distribution over the test dataset can be explained by a
lower level of solar activity and ionospheric disturbance in 2017-2018
compared with 2013-2016. Positive bias means that predicted noise
level is lower than observed one. This bias is
caused by using minimal noise level prediction instead of predicting
the actual noise level. 

When using the noise level for investigations of ionospheric absorption,
the space weather effects usually cause lowering noise level
\cite{BERNGARDT_2018,Bland_2018,Bland_2019,Berngadt_2019}, so using
the minimal noise level prediction makes the absorption identification
technique less sensitive to random variations of the noise level.

\paragraph{Fine prediction model of the noise level}

The rough prediction model described above predicts minimal noise level
using minimal noise levels measured in previous days. Using this model allow
us to identify the periods when noise level is lower than expected
minimal value. 

To predict the most expected noise level one requires an unbiased
forecast, produced a zero mean error between experimental observations
and forecast values. 
To build such an unbiased fine prediction of the noise, we used the
rough noise model (described above) scaled to provide unbiased prediction.

The predicted value of the average noise level $f_{m,0}(t+\Delta T)$
with prediction time $\Delta T$ ahead the current moment $t$ in
this approach can be presented as:

\begin{equation}
f_{m,0}(t+\Delta T)=M(t,\Delta T)f_{m,min}(t+\Delta T)
\label{eq:regrM}
\end{equation}
where $M(t,\Delta T)$ - some scaling factor. Since the experimental
noise level $f_{e}(t)$ in this approach is proportional to the model
one $f_{m,min}(t)$, the task of constructing an unbiased model is
reduced to finding the optimal scaling function $M(t,\Delta T)$. 

In proposed approach $M(t,\Delta T)$ is calculated as the weighted
average of the ratios of the experimental noise level to the rough
model of minimal noise level over a sufficiently long period. For calculations
we use 12 days period with temporal resolution 15 minutes to predict
noise level 12 hours ahead:

\begin{equation}
M(t,12h)=\sum_{\tau=0}^{12\,days}Q(\tau)\frac{f_{e}(t-\tau)}{f_{m,min}(t-\tau)}
\label{eq:regrQ}
\end{equation}
where $Q(\tau)$ are regression coefficients to be found by
fitting the experimental data $f_e(t;f_{sound},B_m)$ by the model Eq.(\ref{eq:regrM},\ref{eq:regrQ}):

\begin{equation}
\varOmega_{2}(f_{sound},B_{m})=\int\left(f_{m,0}(t;f_{sound},B_{m})-f_{e}(t;f_{sound},B_{m})\right)^{2}dt=min
\label{eq:Regr_criteria-1}
\end{equation}

The 12 hours prediction interval is chosen as equidistant
both from the current moment supported by the measurements $f_{e}(t)$
and from the 24 hour forecast value supported by the optimal prediction
of the minimal noise level $f_{m,min}(t+24h)$ . We expect that the
quality of the forecast at the moment $t+12h$ is the worst possible. 
Therefore, using the optimal weighting coefficients $Q(\tau)$
for this very forecast interval should improve the prediction accuracy.

The values of the regression coefficients $Q(\tau)$ are calculated
based on the condition Eq.(\ref{eq:Regr_criteria-1}) over the test dataset (2013-2016). The variability of
these coefficients from $\tau$, beam number $B_m$ and frequency $f_0$ is very
high, which does not allow them to be analyzed directly. Therefore,
we analyze the integral of the regression coefficients:

\begin{equation}
I(T)=\frac{1}{\sqrt{T}}\int_{0}^{T}Q(\tau)d\tau\label{eq:regrI}
\end{equation}

also averaged over frequencies $f_0$ and beams $B_m$.

The position of the maximum of the function $I(T)$ corresponds to
the characteristic time at which the coefficients of the regression model 
Eq.(\ref{eq:regrM},\ref{eq:regrQ}) are effective. The normalization of the integral by $\sqrt{T}^{-1}$
is used for compensating the effect of 'random walk process' on the
estimation of this characteristic time.

\begin{figure}
\includegraphics[scale=0.7]{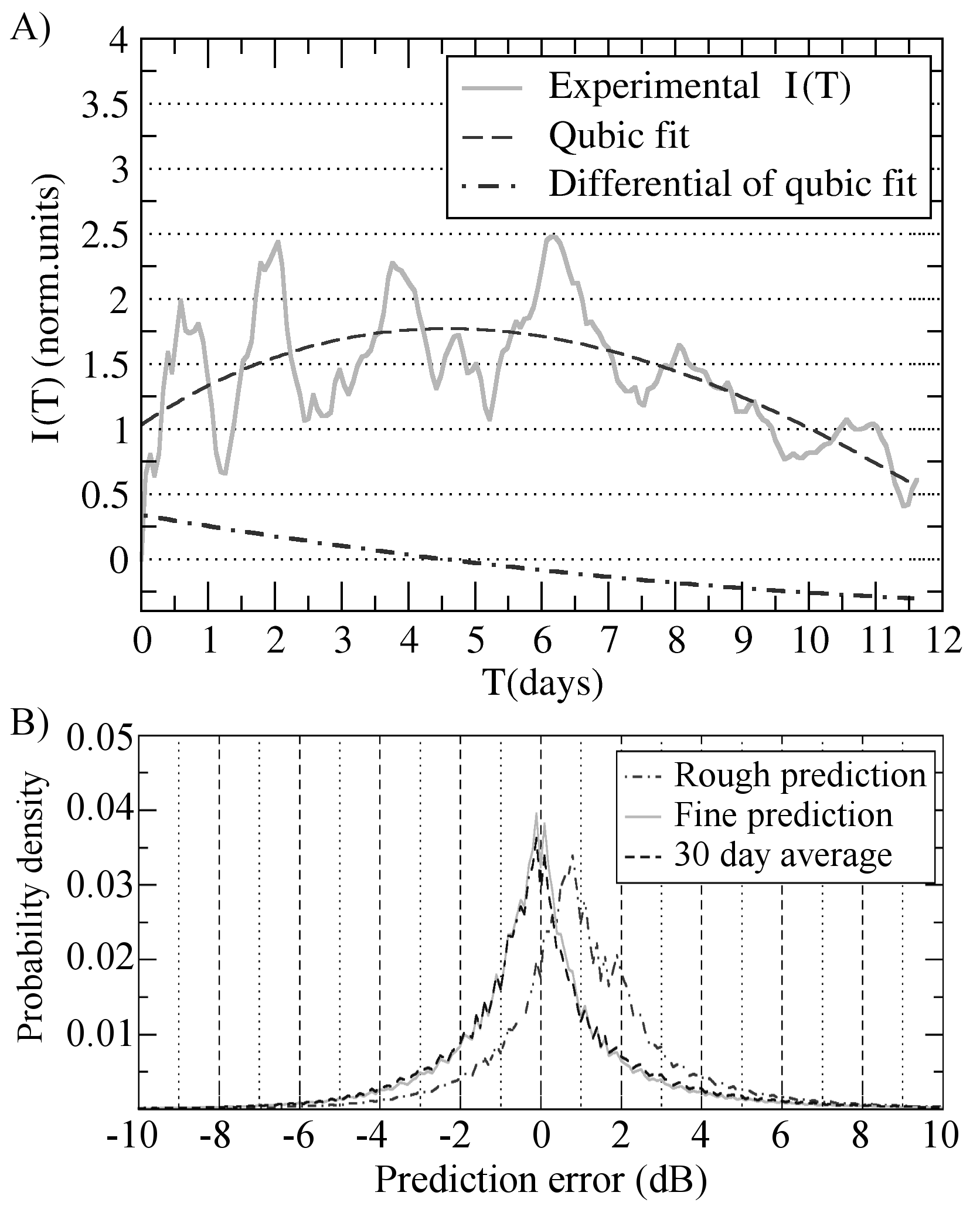}
\caption{A) Dependence of $I(T)$  (Eq.(\ref{eq:regrI}))
on the delay $T$. The solid line is the experimental values of $I(T)$,
the dashed line is its approximation by a smooth curve. The dot-dashed
line corresponds to the expected average behavior of the regression coefficients
$Q(\tau)$. B) The distributions of noise prediction errors for 6 hours
ahead the current time after using 3 different techniques - rough prediction
(dot-dashed line), 30 day average (dashed line) and fine prediction
(solid line).}
\label{fig:Fig3}
\end{figure}

Fig.\ref{fig:Fig3}A shows the behavior of $I(T)$ (Eq.(\ref{eq:regrI}), gray solid line) and its
approximation by smooth cubic curve (dashed line). 

In Fig.\ref{fig:Fig3}A the dot-dashed line shows the expected model
of the regression coefficients $Q(T)$, obtained by differentiating
the smooth polynomial approximation of integral Eq.(\ref{eq:regrI}) (dashed
line). From Fig.\ref{fig:Fig3}A one can see that the
main contribution to the regression prediction of the average noise
level is given by previous 5 days, and their contribution decreases
with from the most recent to the oldest days, and, in the first approximation,
linearly. Thus, we use the simple analytical model of regression
coefficients $Q(\tau)$:

\begin{equation}
Q(\tau)\approx\left\{ \begin{array}{l}
2\frac{T_{5d}-\tau}{T_{5d}};\tau<T_{5d}\\
0;\tau\geq T_{5d}
\end{array}\right.\label{eq:Q_analytic}
\end{equation}
where $T_{5d}$ is the optimal duration of regression, equal to 5
days.

By substituting Eq.(\ref{eq:Q_analytic}) into Eq.(\ref{eq:regrM},\ref{eq:regrQ})
we obtain the fine prediction model in the final form:

\begin{equation}
f_{m,0}(t+\Delta T)=\sum_{\tau=0}^{T_{5d}}\left\{ 2\frac{T_{5d}-\tau}{T_{5d}}\cdot\frac{f_{e}(t-\tau)}{f_{m,min}(t-\tau)}\right\} f_{m,min}(t+\Delta T)\label{eq:regrM-exact}
\end{equation}

The constructed fine prediction model $f_{m,0}(t)$ uses rough prediction
of the minimal noise level $f_{m,min}(t)$ obtained over previous
28 days and scale it using the previous 5 days 
data. So for functioning the fine model prediction requires 33 days of the historical observations. 

Fig.\ref{fig:Fig3}B shows the distribution of errors for various
forecast techniques based on observations by the EKB ISTP SB RAS radar
during 2013-2018. The figure shows a comparison of the 30-day average
forecast (used in riometer measurements \cite{Heisler_1967}, as well as in some radar measurements \cite{Bland_2018},
dashed line), the rough forecast of the minimal noise level described
above (Eq.(\ref{eq:model_daily},\ref{eq:RnModel}), dot-dashed line) and
the fine forecast, described above (Eq.(\ref{eq:regrM-exact}), solid
line). It can be seen from the Fig.\ref{fig:Fig3}B that the
 30-day average forecast has the same error
distribution as fine model has. This allows one to effectively use both forecasts
for predicting actual noise level. The rough prediction model
$f_{m,min}(t)$ is biased, and regularly underestimates the forecast
value $f_{m,min}(t)$ relative to the experimental one $f_{e}(t)$.
This bias approximately corresponds to single RMS error of using $f_{m,0}(t)$.

Thus, we can use $f_{m,min}(t)$ for detecting absorption periods,
as the following condition:

\begin{equation}
f_{e}(t)<f_{m,min}(t)
\label{eq:algoEq}
\end{equation}

The expected error of this detection technique can be determined from
the distribution of errors (fig.\ref{fig:Fig3}B) and is about
25\%. To increase the accuracy of the detection
we analyze each beam-frequency channel independently: for each frequency
range $f_{sound}$ and each beam $B_{m}$. We select only the events
that are simultaneously observed (i.e. lower than corresponding threshold
level Eq.(\ref{eq:algoEq})) independently in five neighboring beams $B_{m}$ and in two
independent frequencies $f_{sound}$. Using this approach should significantly decrease
the detection error.

After detecting the absorption periods the exact
value of the current absorption $A(t)$ is calculated from the fine prediction
model 6 hours ahead based on 6 hours old measurements as following:

\begin{equation}
A(t)=\left\{ \begin{array}{c}
f_{e}(t)-f_{m,0}(t)[dB],\,f_{e}(t)<f_{m,0}(t)\\
0,\,f_{e}(t)>f_{m,0}(t)
\end{array}\right.\label{eq:algoEq-1}
\end{equation}

\section{Statistical characteristics of detected absorption}

\begin{figure}
\includegraphics[scale=0.5]{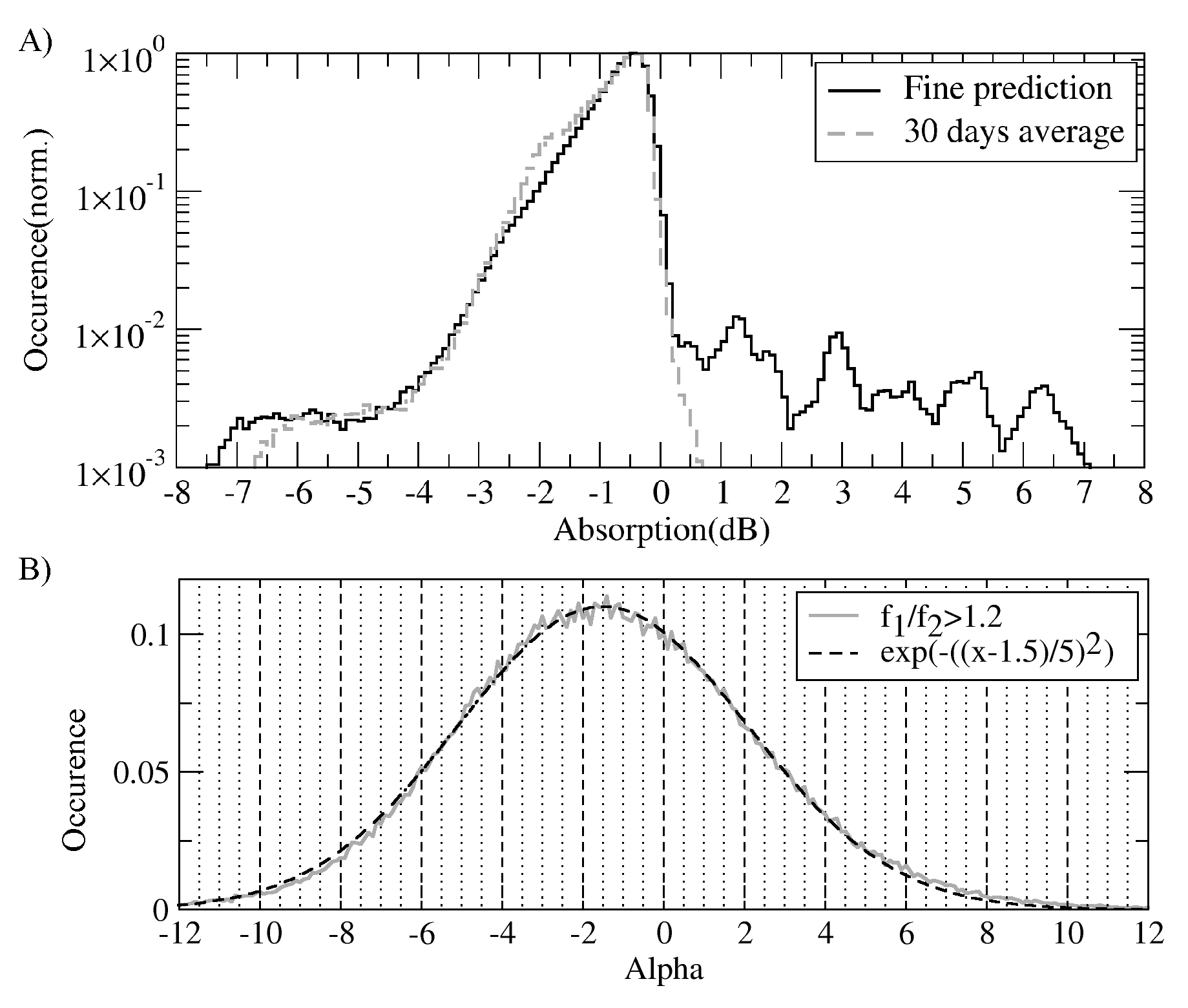}
\caption{A) The distribution of absorption calulated using fine forecast 
and 30 day average;
B) The distribution of the exponential absorption dependence $\alpha$
calculated over the absorption cases with the difference in the sounding
frequencies $f_{1}/f_{2}>1.2$ (orange line) according to the formula
Eq.(\ref{eq:AlphaCalc}). The dashed line shows the approximation of
this distribution by the biased normal distribution.}
\label{fig:Fig4}
\end{figure}

The algorithm for automatic detection of absorption events
and calculation of absorption intensity is used for statistical
studies of spatio-temporal dynamics and frequency dependence of the
detected absorption events.
In Fig.\ref{fig:Fig4}A shown the comparison of absorption distributions 
using 
two prediction models - fine forecast (Fig.\ref{fig:Fig4}A, solid line, Eq.(\ref{eq:algoEq})) and 30-day average value 
(Fig.\ref{fig:Fig4}A, dashed line). Absorption events are selected by Eq.(\ref{eq:algoEq}). The comparison shows that both 
predictions produces nearly the same absorption distributions, 
and both techniques can be used to calculate the absorption.

\paragraph{Frequency dependence of the detected absorption}

Traditionally it is believed that the absorption of radio waves vertically
propagating in the ionosphere has a power-law frequency dependence
\cite{Hargreaves_2010,Schumer2010,DRAP,DRAP2,Bland_2018,Berngadt_2019,Bland_2019}:

\[
A[dB]=B\cdot f_{sound}^{\alpha}
\]

Therefore, one of the possible approaches for verification of the
developed technique is the analysis of frequency dependence of absorption
based on large dataset. It has been evaluated in many papers, and
usually $\alpha\approx-1.3\div-2.0$ \cite{Hargreaves_2010,Schumer2010,DRAP,DRAP2,Bland_2018,Berngadt_2019,Bland_2018}.

By analogy with \cite{Berngadt_2019}, to identify the power-law index
and to increase the accuracy of its estimates, we select all the available
observations of absorption events detected by EKB radar in 2013-2018
and transform them to equivalent vertical absorption (using algorithm
described in \cite{Berngadt_2019}). To improve the accuracy we analyze
only the cases when the absorption was detected simultaneously at
two frequencies separated by more than 1.2 times. The frequency dependence
$\alpha$ in each of the observations is determined as:

\begin{equation}
\alpha=\frac{log(\frac{A_{1}[dB]sin(E_{1})}{A_{2}[dB]\cdot sin(E_{2})})}{log(f_{1}[MHz]/f_{2}[MHz])}\label{eq:AlphaCalc}
\end{equation}
where $A_{1},A_{2},f_{1},f_{2}$ are the logarithms of noise absorption
at two frequencies, and these frequencies, respectively, $E_{1},E_{2}$
are the effective noise elevation angles calculated by algorithm presented in \cite{Berngadt_2019}.

The Fig.\ref{fig:Fig4}B shows the distribution of $\alpha$,
calculated according to Eq.(\ref{eq:AlphaCalc}) over more than 235 thousands of
individual observations and the result of optimal fitting of this
distribution by the normal distribution with a dispersion 5.0 and an
average of -1.5. It can be seen from the figure that the 
frequency dependence with $\alpha=-1.5$ satisfactorily fits the
experimental data and is in good agreement with the results
 obtained in papers \cite{Berngadt_2019,DRAP,DRAP2,Schumer2010,Rogers_2015}.

Thus, the simultaneous observation of radio noise
absorption at two spaced frequencies ($f_{1}/f_{2}>1.2$) 
can be interpreted
as signal absorption in the ionosphere. This makes it possible to
use the obtained noise prediction model and the detection algorithm
for monitoring absorption in the ionosphere by coherent decameter
radars.

The resulting absorption calculation algorithm is the following:
\begin{itemize}
\item At two frequencies and 5 neighboring beams, rough (Eq.(\ref{eq:model_daily},\ref{eq:RnModel}))
and fine (Eq.(\ref{eq:regrM-exact})) forecasts of the noise level are
separately made;
\item If in all the cases simultaneously the measured noise 
is below its rough forecast (Eq.(\ref{eq:algoEq})) the moment
 is considered as an absorption event;
\item In each of the frequency channel, the difference between the experimentally measured
noise level and its fine forecast is calculated and reduced to the
equivalent vertical absorption at 10MHz frequency according to:

\begin{equation}
A(t)[dB]=\left\{ \begin{array}{c}
\left(f_{e}(t)-f_{m,0}(t)\right)sin(E(t))\left(\frac{f_{sound}(t)[MHz]}{10MHz}\right)^{1.5},\,f_{e}(t)<f_{m,0}(t)\\
0,\,f_{e}(t)>f_{m,0}(t)
\end{array}\right.
\label{eq:FinalAlgo}
\end{equation}

 where $f_{sound}(t)$ - sounding frequency and $E(t)$ is elevation
calculated by method, described in \cite{Berngadt_2019}.
\item The resulting $A(t)$ is averaged over the 5 neighbor beams and 2 operational frequencies.
\end{itemize}

~

\paragraph{Spatio-temporal dynamics of detected absorption regions}

\begin{figure}
\includegraphics[scale=0.6]{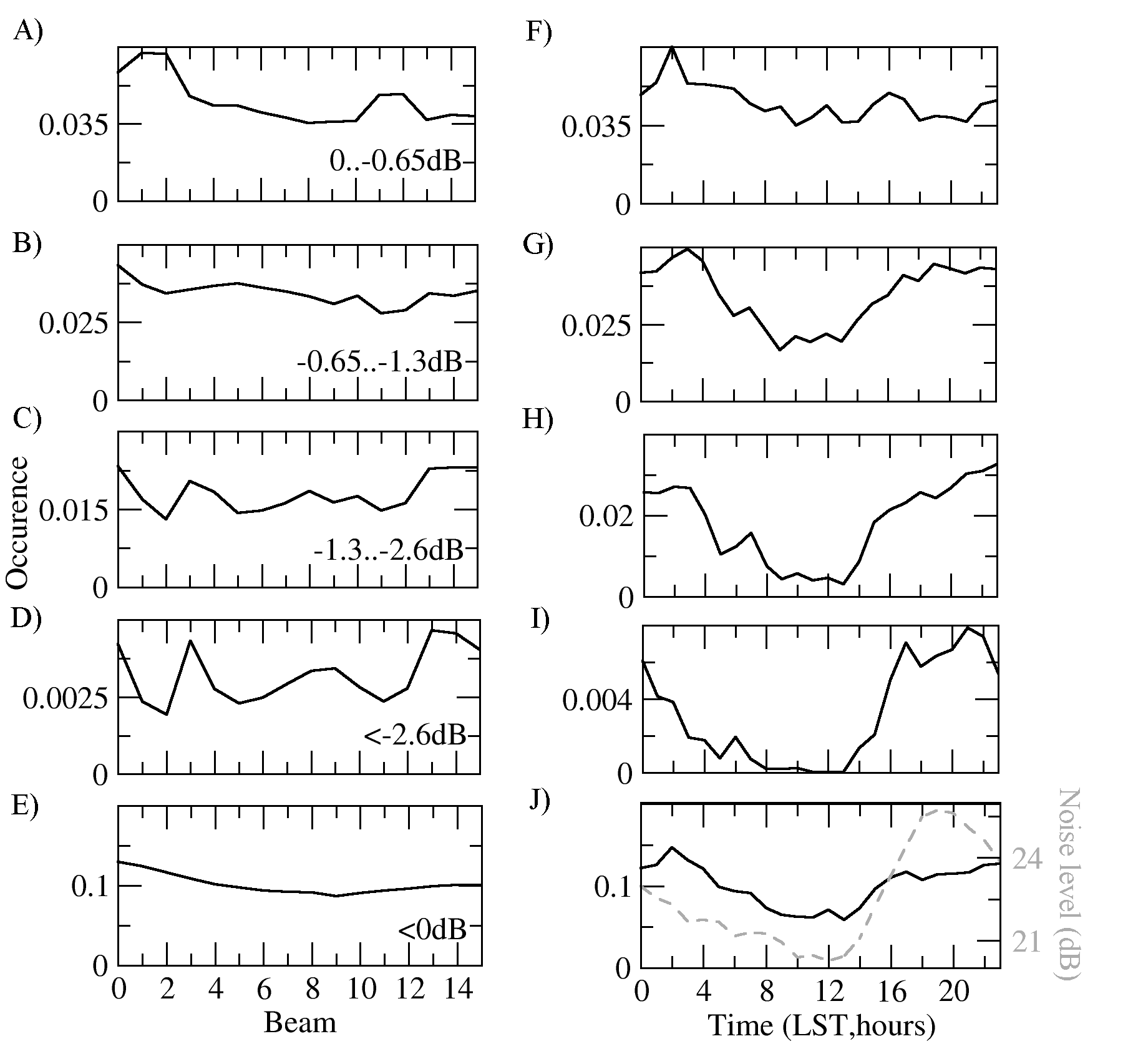}
\caption{
Spatio-temporal dynamics of absorption events with amplitudes 0..-0.65dB (A,F), 
-0.65..-1.3dB(B,G), -1.3..-2.6dB(C,H), <2.6dB(D,I), 
as well as a complete set of absorption events (E,J).
A-E) Absorption events dependence on beam number;
F-J) Absorption events dependence on local solar time;
Gray dashed line at E) corresponds to the average noise level 
as a function of local solar time.}
\label{fig:Fig5}
\end{figure}

Fig.\ref{fig:Fig5} shows the statistics of spatio-temporal
absorption characteristics over 2013-2018 radar dataset.

It can be seen from Fig.\ref{fig:Fig4}A that the most probable absorption 
level is about -0.65 dB. We have analyzed the spatio-temporal dynamics of 
absorption in the ranges 0..-0.65dB, -0.65..-1.3dB, -1.3..-2.6dB, <2.6dB, 
as well as a complete set of absorption events.
The azimuthal dependence of the frequency of observation of absorption events 
Fig.\ref{fig:Fig5}A-E and the dependence of the frequency of observation of 
absorption events on LST (Fig.\ref{fig:Fig5}F-J) are studied.
It can be seen from the Fig.\ref{fig:Fig5} that the absorption with weak and 
high amplitudes differ significantly.
Cases of weak absorption (0..-1.3dB) are more often observed near the northern 
direction and are weakly dependent on local time. Strong events are practically 
independent of the beam number, and their daily dynamics (Fig.\ref{fig:Fig5}H) 
correlates well with the daily dynamics of noise (Fig.\ref{fig:Fig5}J, dashed line).

\begin{figure}
\includegraphics[scale=0.1]{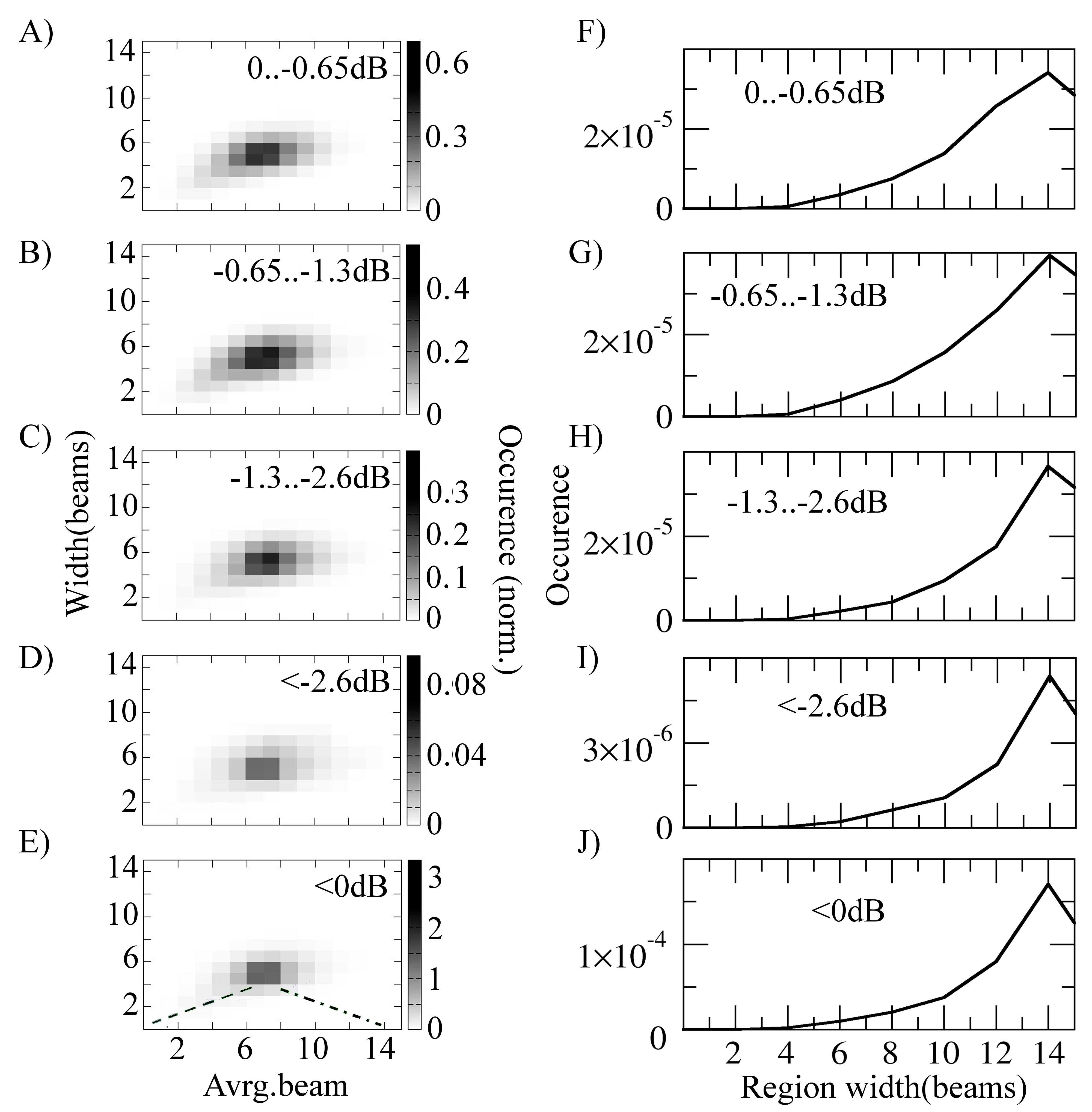}
\caption{
Spatial distributions of absorption events with amplitudes 0..-0.65dB (A,F), 
-0.65..-1.3dB(B,G), -1.3..-2.6dB(C,H), <2.6dB(D,I), 
as well as a complete set of absorption events (E,J).
A-E) Distribution of absorption region locations and sizes (in number of beams);
F-J) Distribution of average absorption region sizes (number of beams)}
\label{fig:Fig6}
\end{figure}

In Fig.\ref{fig:Fig6}A-E shown the distributions of the average
beam and the RMS of the sizes of the regions in which an absorption
is observed with different intensity levels described above. To interpret the distribution of Fig.\ref{fig:Fig6}A-E,
the behavior of the average and RMS was simulated for different locations
and sizes of the region. The simple simulation showed that when the
region is located starting from the 0-th beam and changing is size 
from 1 to 15 beams it produces the dependence of RMS on average
beam corresponding to the dashed line in Fig.\ref{fig:Fig6}E.
When the region is located starting from the 15th beam and changing its size from 
1 to 15 beams, it produces the dependence of RMS
on the average beam corresponding to the dot-dashed line in Fig.\ref{fig:Fig6}E.
From this results we can conclude that experimental observations
at low absorption level (Fig.\ref{fig:Fig6}A-B)
correspond to the case when the absorption areas mainly include the
0th beam. The modeling showed that the average beam of the region
in this case is equal to half the size of the region, which allows
us to estimate the frequency of occurrence of absorbing regions with
different sizes. Such an estimate is shown in Fig.\ref{fig:Fig6}F-J.
It can be seen that most frequently the absorbing
region occupies the entire radar field-of-view, and the occurrence
of smaller regions decreases with decreasing size of the region.

Smaller absorption events are frequently observed with smaller spatial sizes
(Fig.\ref{fig:Fig7}F-G), and higher absorption events - with larger spatial 
sizes (Fig.\ref{fig:Fig7}H-J).

From the figure one can conclude that some cases of absorption events with 
small amplitudes (0 ..-1.3dB) have smaller spatial scales and are concentrated 
on northern beams. The observations with high absorption amplitudes are most 
often large-scale.

\begin{figure}
\includegraphics[scale=0.6]{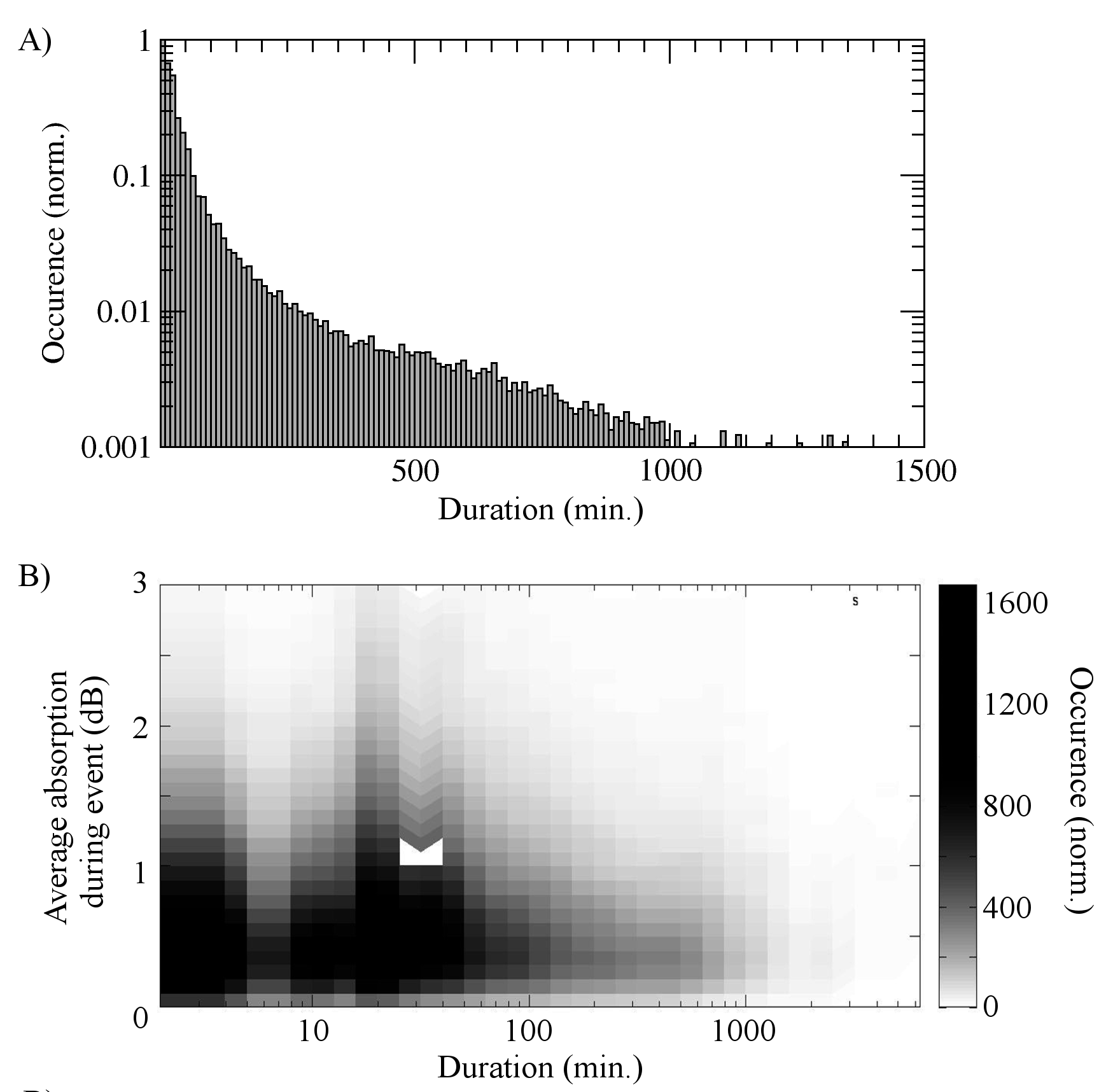}
\caption{
A) Distribution of absorption events by duration;
B) Average absorption amplitudes as function of 
absorption event duration
}
\label{fig:Fig9}
\end{figure}

In Fig.\ref{fig:Fig9}A are shown distributions of absorption events by duration.
As one can see, the observed events usually are shorter than 16 hours, that can 
be related with mid-latitude location of the radar. For high-latitude radars the
durations can be longer.

In Fig.\ref{fig:Fig9}B are shown the average absorption amplitudes over different 
absorption events. As one can see, the most powerful events usually shorter than 
less powerful ones. From the figure ane can see that powerful events up to -2dB 
usually are abserved not longer than 1-1.5 hours.

\section{Case comparison of the prediction algorithms in different cases.}

\paragraph{Short-lived absorption case}

\begin{figure}
\includegraphics[scale=0.5]{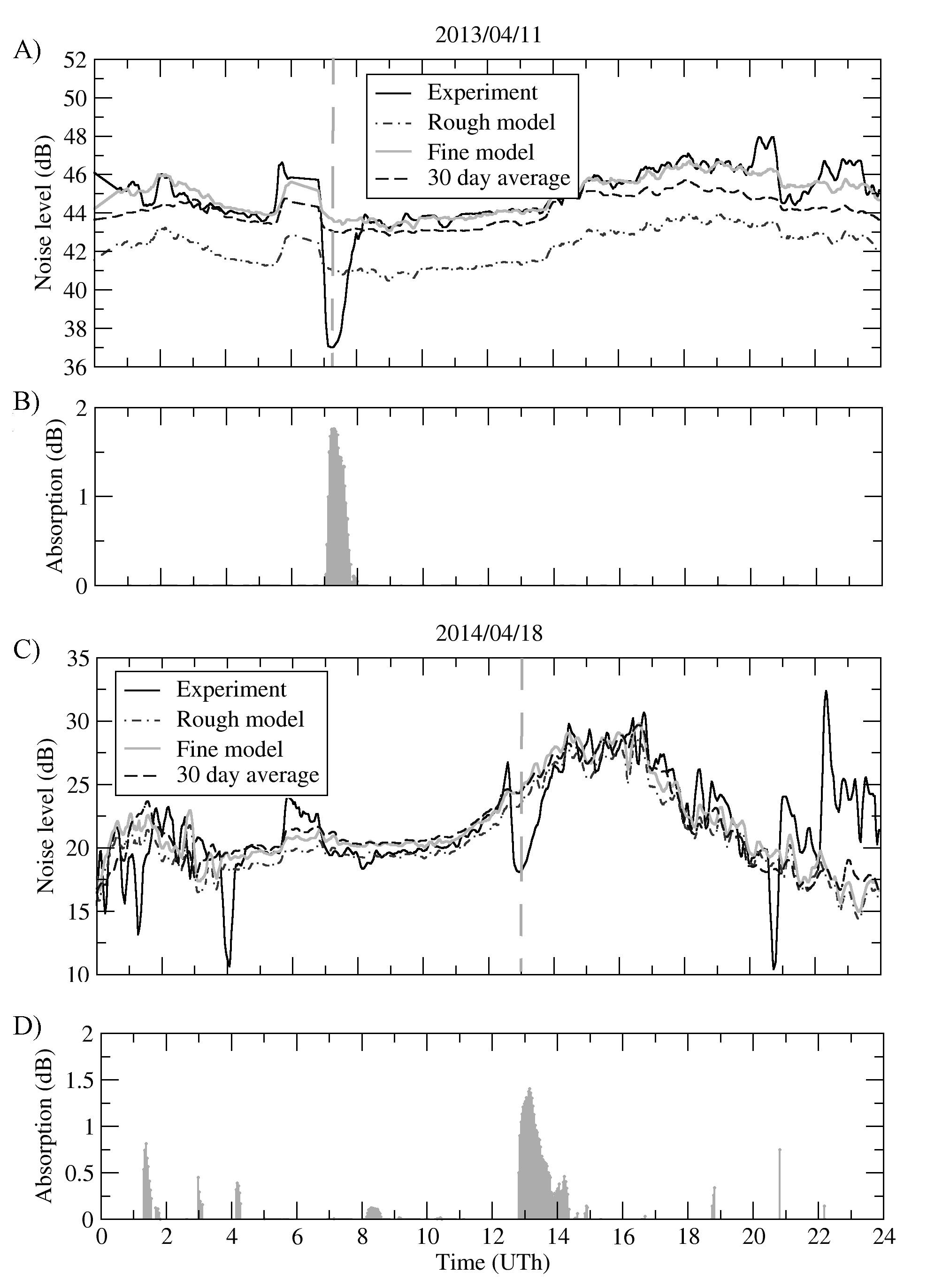}
\caption{The operation of the prediction algorithms during two solar X-ray
flares - 04/11/2013 (A-B) and 18/04/2014 (C-D) (the moments of the
flares are marked by vertical dashed lines). A,C) - noise measurements
and different predictions; B,D) - absorption level calculated by the
algorithm.}
\label{fig:Fig7}
\end{figure}

Fig.\ref{fig:Fig7}A,C shows an example of different algorithms operation 
during two solar X-ray flares:
rough and fine forecasts as well as traditional 30 day average forecast. 
Fig.\ref{fig:Fig7}B,D shows the corresponding
examples of the calculated absorption level, reduced to 10MHz frequency
and to the vertical propagation of a radio wave. It can be seen from
the figures that a rough forecast is often below the noise level
and its forecasts (fine and 30-day average), which leads to
the fact that the use of an fine forecast and a 30-day averaging to
determine the absorption in some cases will give fairly close results,
which is observed experimentally (Fig.\ref{fig:Fig3}B).

As one can see from \ref{fig:Fig7}A, in some cases the fine
prediction algorithm predict noise level better than traditional 30-day
forecast. From \ref{fig:Fig7}C one can also see that the rough
prediction is expectedly about 1 dB lower than actual noise level,
and therefore can be used for detecting the absorption cases. At high
latitudes, for which riometric and radar \cite{Bland_2019} absorption
are large, using the traditional 30 day average forecast instead of the fine prediction
looks equivalent. At the middle latitudes the absorption is smaller
and one should use more accurate fine forecast.

\paragraph{Long-lived absorption case}

\begin{figure}
\includegraphics[scale=0.45]{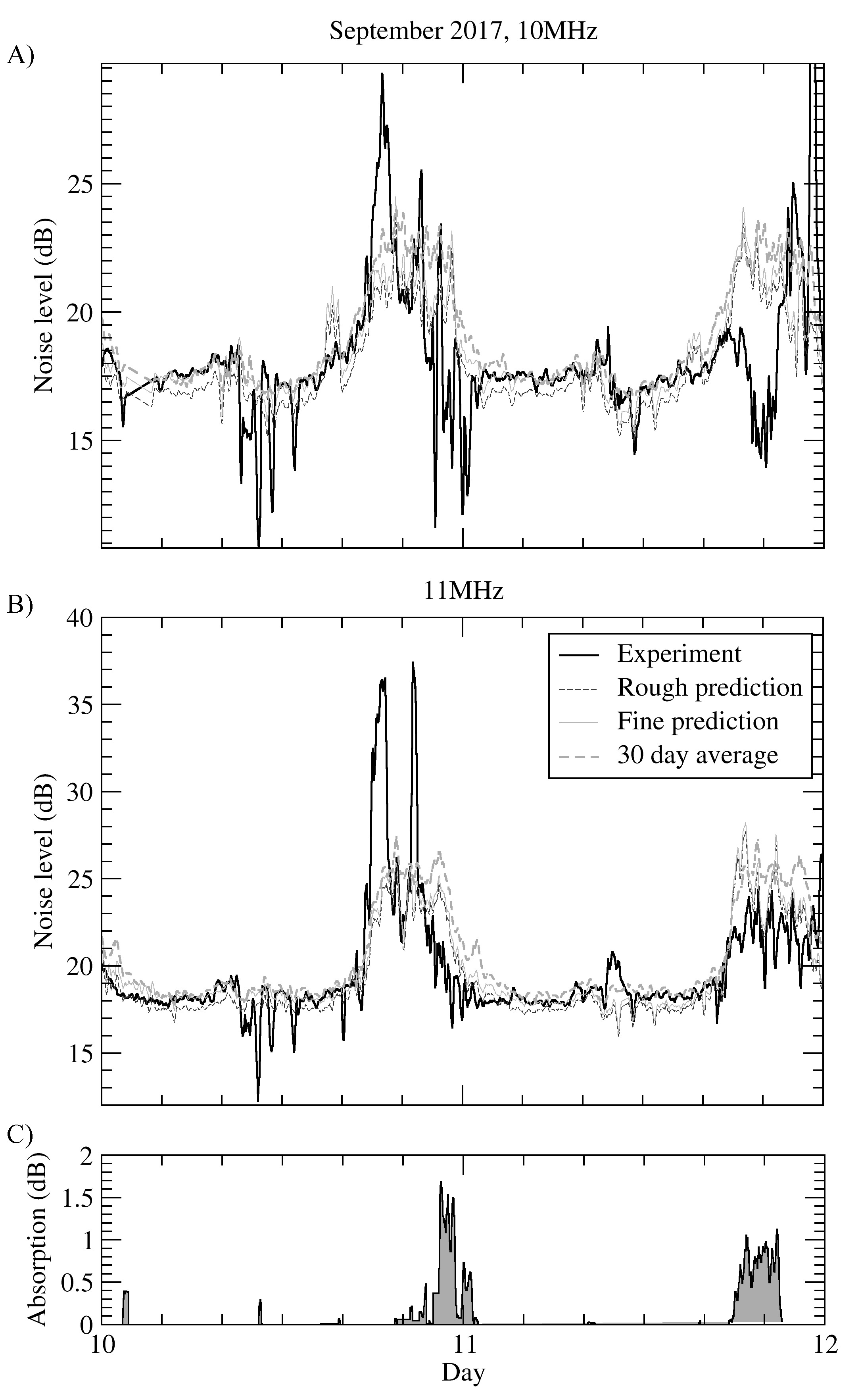}
\caption{An example of the operation of noise prediction algorithms during
the September 10-11, 2017 event. A) observations at 10MHz sounding
frequency B) is observations at 11 MHz frequency. C) is absorption
level detected by final algorithm Eq.(\ref{eq:FinalAlgo}).}
\label{fig:Fig8}
\end{figure}

Fig.\ref{fig:Fig8} shows prediction algorithms during the September
10-11, 2017 event related with Coronal mass ejection event\cite{Bland_2018}.
The Fig.\ref{fig:Fig8}A-B shows the noise
measurements and their prediction at 10 and 11MHz. As one can see
from Fig.\ref{fig:Fig8}B, the traditional 30-day average algorithm
(dashed line) in this case looks working better than fine prediction
algorithm (thin solid line). This can be related with using 5 days adaptation
for fine prediction, that does not work well in case of long-lasting
absorption events. So the fine prediction can be used only in short-lasting events,
for example, during X-ray solar flares \cite{BERNGARDT_2018,Berngadt_2019}.
 Fig.\ref{fig:Fig8}C shows the corresponding
example of the calculated absorption level, reduced to 10MHz frequency
and to the vertical propagation of a radio wave. It also takes into
account observations at both frequencies and at five neighboring beams.

As shows the qualitative case analysis above, the traditional 30-day
averaging can be used for prediction noise level in long-lasting absorption
cases, for fine prediction of noise level in short-lasting events
it is recommended to use fine prediction algorithm. Rough prediction
algorithm can be used for detection the absorption in both cases.

\section{Conclusion}

A statistical analysis of the dynamics of noise variations on a coherent decameter
radar EKB ISTP SB RAS, related to the radiowaves absorption in the lower part of
the ionosphere, was carried out for the first time. To carry out the
analysis, a method for predicting the minimal noise level based on
the optimal autoregression was implemented. The autoregression
is based in the minimal noise measurements over the previous
28 days, and used to forecast the
minimal noise level for 1 day ahead the most accurately. An analytical
formula (Eq.(\ref{eq:RnModel})) is given for the obtained coefficients.

Based on the optimal forecast of the minimal noise level, a method
is proposed for the optimal (fine) forecast of the noise level based
on the calibration (scaling) of the predicted minimal noise level. The analysis
of the complete data set 2013-2018 showed that the optimal forecast
can be build by using calibration over the previous
5 days. The accuracy of the constructed fine forecast is compared
with accuracy of 30 days average noise value. The comparison showed that the
distribution of prediction errors is almost
identical (Fig.\ref{fig:Fig3}B) for these two methods.

 A case analysis of several
events showed that in the case of short-term absorption cases, the
fine prediction model can be more accurate than the traditional riometric
algorithm (Fig.\ref{fig:Fig7}A), and during long-term absorption
cases, the traditional riometric algorithm scan be more accurate (Fig.\ref{fig:Fig8}
A-B). So in the cases of analysis high-latitude data with frequent absorption
the riometric algorithm looks preferable, and in the case of mid-latitude
data and short-living events - fine prediction algorithm looks preferable.

Based on the analysis of September 10-11, 2017 case, it was shown that
at different frequencies the shape of the absorption effect may vary
(Fig.\ref{fig:Fig8} A-B), which leads to the need to use measurements
at several frequencies to more reliably detect the cases of absorption.

As a result, a two-frequency detection technique is proposed for
detecting the absorption cases, based on simultaneously
detecting a decrease in noise level below the rough minimal noise forecast
at two independent frequencies and five neighboring radar beams. 
The absorption is estimated by 
calculating the difference in the measured noise level relative to
the fine noise level forecast and averaging the effect over sounding frequencies and neighbor beams, taking into
account the frequency and elevation angle dependencies of the absorption
and reduction the absorption value to 10MHz frequency in analogy to
\cite{Berngadt_2019}.

Statistical analysis of two-frequency experiments in the period 2013-2018
is made. The analysis shows the exponential frequency dependence of
the absorption with -1.5 power coefficient (Fig.\ref{fig:Fig4}B).
This value corresponds well with the data of other observations and models
\cite{Hargreaves_2010,Schumer2010,DRAP,DRAP2,Berngadt_2019}
and allows to use the proposed model for detecting and estimating
the absorption level over the large noise dataset obtained at
EKB ISTP SB RAS radar in 2013-2018.

Based on a large amount of statistical data (2013-2018), it was shown
that the average daily diurnal variation of the vertical absorption
level at 10 MHz does not depend on local time for abosrption events with 
smaller amplitude (0..-1.3dB) and looks close to the variations of the average durinal
variation of the noise level (Fig.\ref{fig:Fig5}C,E) at high amplitudes.

Based on the statistical analysis of the spatial structure of the
absorbing regions over 2013-2018, it was shown that a significant
part of the absorption events with small amplitude (0..-1.3dB) corresponds 
to high-latitude absorption(Fig.\ref{fig:Fig6}F-G), and high amplitude 
absorption events (stonger than -1.3dB) corresponds to large-scale events comparable
in size to the radar field-of-view (Fig.\ref{fig:Fig6}H-J)
- about 50 degrees in azimuth. This allows to suggest that observed strong absorption events 
at nighttime can be significantly affected by radiowave 
propagation effects, requires more correct elevation angle calculations 
(significantly changing calculated absorption level)
and should be interpreted as absorption very carefully.

It is important to note that by definition the presented technique can not be used for 
investigations of regular absorption variations with 24 hours periodicity, and is useful 
for investigating irregular, preferably short-living events.

\section*{Acknowledgements}
EKB ISTP SB RAS facility from Angara Center for Common Use 
of scientific equipment (http://ckp-rf.ru/ckp/3056/) is operated under budgetary funding of Basic Research program II.12. 
The data of EKB ISTP SB RAS
radar are available at ISTP SB RAS (http://sdrus.iszf.irk.ru/ekb/page\_
example/simple).
The work is supported by RFBR grant 18-05-00539a.

\section*{References}

\end{document}